\RequirePackage[mathlines]{lineno}
\documentclass[amsmath,amssymb,twocolumn,prd,floatfix,showpacs, nofootinbib]{revtex4-1}
\usepackage{tabularx}  
\usepackage{bm, color} 
\usepackage{overpic,subfigure} 
\usepackage{multirow}
\usepackage{array}
\usepackage{dcolumn} 
\usepackage[symbol]{footmisc}
\usepackage{epstopdf}
\usepackage{ulem}
\RequirePackage{xspace}

\newcommand{\gev}{\ensuremath{\mathrm{\,Ge\kern -0.1em V}}\xspace}
\newcommand{\mev}{\ensuremath{\mathrm{\,Me\kern -0.1em V}}\xspace}
\newcommand{\mevcc}{\ensuremath{{\mathrm{\,Me\kern -0.1em V\!/}c^2}}\xspace}

\def\chic#1{\ensuremath{\chi_{c#1}}\xspace} 

\def\fz#1       {\ensuremath{f_0({#1})}\xspace}

\usepackage{lineno}
\usepackage{hyperref}
\hypersetup{hypertex=true,
	        colorlinks = true,
            linkcolor = blue,
            urlcolor = blue,
            citecolor = blue}
            
\begin{document}

\title{\boldmath Observation of $\chi_{cJ}\to p \bar p K^0_S K^- \pi^+ + c.c.$ }

\author{
	M.~Ablikim$^{1}$, M.~N.~Achasov$^{4,c}$, P.~Adlarson$^{76}$, O.~Afedulidis$^{3}$, X.~C.~Ai$^{81}$, R.~Aliberti$^{35}$, A.~Amoroso$^{75A,75C}$, Y.~Bai$^{57}$, O.~Bakina$^{36}$, I.~Balossino$^{29A}$
% ...
	\\
	\vspace{0.2cm}
	(BESIII Collaboration)\\
	\vspace{0.2cm} {\it
		$^{1}$ Institute of High Energy Physics, Beijing 100049, People's Republic of China\\
		$^{2}$ Beihang University, Beijing 100191, People's Republic of China\\
		$^{3}$ Bochum  Ruhr-University, D-44780 Bochum, Germany\\
		$^{4}$ Budker Institute of Nuclear Physics SB RAS (BINP), Novosibirsk 630090, Russia\\
		$^{5}$ Carnegie Mellon University, Pittsburgh, Pennsylvania 15213, USA\\
		$^{6}$ Central China Normal University, Wuhan 430079, People's Republic of China\\
		$^{7}$ Central South University, Changsha 410083, People's Republic of China\\
		$^{8}$ China Center of Advanced Science and Technology, Beijing 100190, People's Republic of China\\
		$^{9}$ China University of Geosciences, Wuhan 430074, People's Republic of China\\
		$^{10}$ Chung-Ang University, Seoul, 06974, Republic of Korea\\
		$^{11}$ COMSATS University Islamabad, Lahore Campus, Defence Road, Off Raiwind Road, 54000 Lahore, Pakistan\\
		$^{12}$ Fudan University, Shanghai 200433, People's Republic of China\\
		$^{13}$ GSI Helmholtzcentre for Heavy Ion Research GmbH, D-64291 Darmstadt, Germany\\
		$^{14}$ Guangxi Normal University, Guilin 541004, People's Republic of China\\
		$^{15}$ Guangxi University, Nanning 530004, People's Republic of China\\
		$^{16}$ Hangzhou Normal University, Hangzhou 310036, People's Republic of China\\
		$^{17}$ Hebei University, Baoding 071002, People's Republic of China\\
		$^{18}$ Helmholtz Institute Mainz, Staudinger Weg 18, D-55099 Mainz, Germany\\
		$^{19}$ Henan Normal University, Xinxiang 453007, People's Republic of China\\
		$^{20}$ Henan University, Kaifeng 475004, People's Republic of China\\
		$^{21}$ Henan University of Science and Technology, Luoyang 471003, People's Republic of China\\
		$^{22}$ Henan University of Technology, Zhengzhou 450001, People's Republic of China\\
		$^{23}$ Huangshan College, Huangshan  245000, People's Republic of China\\
		$^{24}$ Hunan Normal University, Changsha 410081, People's Republic of China\\
		$^{25}$ Hunan University, Changsha 410082, People's Republic of China\\
		$^{26}$ Indian Institute of Technology Madras, Chennai 600036, India\\
		$^{27}$ Indiana University, Bloomington, Indiana 47405, USA\\
		$^{28}$ INFN Laboratori Nazionali di Frascati , (A)INFN Laboratori Nazionali di Frascati, I-00044, Frascati, Italy; (B)INFN Sezione di  Perugia, I-06100, Perugia, Italy; (C)University of Perugia, I-06100, Perugia, Italy\\
		$^{29}$ INFN Sezione di Ferrara, (A)INFN Sezione di Ferrara, I-44122, Ferrara, Italy; (B)University of Ferrara,  I-44122, Ferrara, Italy\\
		$^{30}$ Inner Mongolia University, Hohhot 010021, People's Republic of China\\
		$^{31}$ Institute of Modern Physics, Lanzhou 730000, People's Republic of China\\
		$^{32}$ Institute of Physics and Technology, Peace Avenue 54B, Ulaanbaatar 13330, Mongolia\\
		$^{33}$ Instituto de Alta Investigaci\'on, Universidad de Tarapac\'a, Casilla 7D, Arica 1000000, Chile\\
		$^{34}$ Jilin University, Changchun 130012, People's Republic of China\\
		$^{35}$ Johannes Gutenberg University of Mainz, Johann-Joachim-Becher-Weg 45, D-55099 Mainz, Germany\\
		$^{36}$ Joint Institute for Nuclear Research, 141980 Dubna, Moscow region, Russia\\
		$^{37}$ Justus-Liebig-Universitaet Giessen, II. Physikalisches Institut, Heinrich-Buff-Ring 16, D-35392 Giessen, Germany\\
		$^{38}$ Lanzhou University, Lanzhou 730000, People's Republic of China\\
		$^{39}$ Liaoning Normal University, Dalian 116029, People's Republic of China\\
		$^{40}$ Liaoning University, Shenyang 110036, People's Republic of China\\
		$^{41}$ Nanjing Normal University, Nanjing 210023, People's Republic of China\\
		$^{42}$ Nanjing University, Nanjing 210093, People's Republic of China\\
		$^{43}$ Nankai University, Tianjin 300071, People's Republic of China\\
		$^{44}$ National Centre for Nuclear Research, Warsaw 02-093, Poland\\
		$^{45}$ North China Electric Power University, Beijing 102206, People's Republic of China\\
		$^{46}$ Peking University, Beijing 100871, People's Republic of China\\
		$^{47}$ Qufu Normal University, Qufu 273165, People's Republic of China\\
		$^{48}$ Renmin University of China, Beijing 100872, People's Republic of China\\
		$^{49}$ Shandong Normal University, Jinan 250014, People's Republic of China\\
		$^{50}$ Shandong University, Jinan 250100, People's Republic of China\\
		$^{51}$ Shanghai Jiao Tong University, Shanghai 200240,  People's Republic of China\\
		$^{52}$ Shanxi Normal University, Linfen 041004, People's Republic of China\\
		$^{53}$ Shanxi University, Taiyuan 030006, People's Republic of China\\
		$^{54}$ Sichuan University, Chengdu 610064, People's Republic of China\\
		$^{55}$ Soochow University, Suzhou 215006, People's Republic of China\\
		$^{56}$ South China Normal University, Guangzhou 510006, People's Republic of China\\
		$^{57}$ Southeast University, Nanjing 211100, People's Republic of China\\
		$^{58}$ State Key Laboratory of Particle Detection and Electronics, Beijing 100049, Hefei 230026, People's Republic of China\\
		$^{59}$ Sun Yat-Sen University, Guangzhou 510275, People's Republic of China\\
		$^{60}$ Suranaree University of Technology, University Avenue 111, Nakhon Ratchasima 30000, Thailand\\
		$^{61}$ Tsinghua University, Beijing 100084, People's Republic of China\\
		$^{62}$ Turkish Accelerator Center Particle Factory Group, (A)Istinye University, 34010, Istanbul, Turkey; (B)Near East University, Nicosia, North Cyprus, 99138, Mersin 10, Turkey\\
		$^{63}$ University of Bristol, (A)H H Wills Physics Laboratory; (B)Tyndall Avenue; (C)Bristol; (D)BS8 1TL\\
		$^{64}$ University of Chinese Academy of Sciences, Beijing 100049, People's Republic of China\\
		$^{65}$ University of Groningen, NL-9747 AA Groningen, The Netherlands\\
		$^{66}$ University of Hawaii, Honolulu, Hawaii 96822, USA\\
		$^{67}$ University of Jinan, Jinan 250022, People's Republic of China\\
		$^{68}$ University of Manchester, Oxford Road, Manchester, M13 9PL, United Kingdom\\
		$^{69}$ University of Muenster, Wilhelm-Klemm-Strasse 9, 48149 Muenster, Germany\\
		$^{70}$ University of Oxford, Keble Road, Oxford OX13RH, United Kingdom\\
		$^{71}$ University of Science and Technology Liaoning, Anshan 114051, People's Republic of China\\
		$^{72}$ University of Science and Technology of China, Hefei 230026, People's Republic of China\\
		$^{73}$ University of South China, Hengyang 421001, People's Republic of China\\
		$^{74}$ University of the Punjab, Lahore-54590, Pakistan\\
		$^{75}$ University of Turin and INFN, (A)University of Turin, I-10125, Turin, Italy; (B)University of Eastern Piedmont, I-15121, Alessandria, Italy; (C)INFN, I-10125, Turin, Italy\\
		$^{76}$ Uppsala University, Box 516, SE-75120 Uppsala, Sweden\\
		$^{77}$ Wuhan University, Wuhan 430072, People's Republic of China\\
		$^{78}$ Yantai University, Yantai 264005, People's Republic of China\\
		$^{79}$ Yunnan University, Kunming 650500, People's Republic of China\\
		$^{80}$ Zhejiang University, Hangzhou 310027, People's Republic of China\\
		$^{81}$ Zhengzhou University, Zhengzhou 450001, People's Republic of China\\
}		
		\vspace{0.2cm}
		$^{a}$ Deceased\\
		$^{b}$ Also at the Moscow Institute of Physics and Technology, Moscow 141700, Russia\\
		$^{c}$ Also at the Novosibirsk State University, Novosibirsk, 630090, Russia\\
		$^{d}$ Also at the NRC "Kurchatov Institute", PNPI, 188300, Gatchina, Russia\\
		$^{e}$ Also at Goethe University Frankfurt, 60323 Frankfurt am Main, Germany\\
		$^{f}$ Also at Key Laboratory for Particle Physics, Astrophysics and Cosmology, Ministry of Education; Shanghai Key Laboratory for Particle Physics and Cosmology; Institute of Nuclear and Particle Physics, Shanghai 200240, People's Republic of China\\
		$^{g}$ Also at Key Laboratory of Nuclear Physics and Ion-beam Application (MOE) and Institute of Modern Physics, Fudan University, Shanghai 200443, People's Republic of China\\
		$^{h}$ Also at State Key Laboratory of Nuclear Physics and Technology, Peking University, Beijing 100871, People's Republic of China\\
		$^{i}$ Also at School of Physics and Electronics, Hunan University, Changsha 410082, China\\
		$^{j}$ Also at Guangdong Provincial Key Laboratory of Nuclear Science, Institute of Quantum Matter, South China Normal University, Guangzhou 510006, China\\
		$^{k}$ Also at MOE Frontiers Science Center for Rare Isotopes, Lanzhou University, Lanzhou 730000, People's Republic of China\\
		$^{l}$ Also at Lanzhou Center for Theoretical Physics, Lanzhou University, Lanzhou 730000, People's Republic of China\\
		$^{m}$ Also at the Department of Mathematical Sciences, IBA, Karachi 75270, Pakistan\\
		$^{n}$ Also at Ecole Polytechnique Federale de Lausanne (EPFL), CH-1015 Lausanne, Switzerland\\
		$^{o}$ Also at Helmholtz Institute Mainz, Staudinger Weg 18, D-55099 Mainz, Germany\\
}

\begin{abstract}
  By analyzing $(27.12\pm0.14)\times10^8$ $\psi(3686)$ events collected with the BESIII detector operating at the BEPCII collider, the decays of $\chi_{cJ} \to p \bar{p} K^0_S K^- \pi^+ +c.c.(J=0, 1, 2)$ are observed for the first time with statistical significances greater than $10\sigma$.
  The branching fractions of these decays are determined to be
  		$\mathcal{B}(\chi_{c0}\to p \bar p K^{0}_{S} K^- \pi^+ + c.c.)=(2.61\pm0.27\pm0.32)\times10^{-5},$
  $\mathcal{B}(\chi_{c1}\to p \bar p K^{0}_{S} K^- \pi^+ + c.c.)=(4.16\pm0.24\pm0.46)\times10^{-5},$ and 
  $\mathcal{B}(\chi_{c2}\to p \bar p K^{0}_{S} K^- \pi^+ + c.c.)=(5.63\pm0.28\pm0.46)\times10^{-5}$, respectively.
The processes $\chi_{c1,2} \to \bar{p} \Lambda(1520) K^0_S \pi^{+} + c.c.$ are also observed, with
 statistical significances of 5.7$\sigma$ and 7.0$\sigma$, respectively. Evidence for $\chi_{c0} \to\bar{p} \Lambda(1520) K^0_S \pi^{+} + c.c.$ is found with statistical significances of 3.3$\sigma$ each.
The corresponding branching fractions are determined to be
	$\mathcal{B}(\chi_{c0}\to \bar{p} \Lambda(1520) K^0_S \pi^{+} + c.c.) =(1.61^{+0.68}_{-0.64}\pm0.23)\times10^{-5}$,
$\mathcal{B}(\chi_{c1}\to \bar{p} \Lambda(1520) K^0_S \pi^{+} + c.c.)=(4.06^{+0.80}_{-0.76}\pm0.52)\times10^{-5}$, and 
$\mathcal{B}(\chi_{c2}\to \bar{p} \Lambda(1520) K^0_S \pi^{+} + c.c.)=(4.09^{+0.87}_{-0.84}\pm0.42)\times10^{-5}$.
Here, the first uncertainties are statistical and the second ones are systematic.

\end{abstract}

\maketitle

\section{Introduction}
Experimental studies of charmonium states and their decay properties are important to test Quantum chromodynamics~(QCD) models and QCD based calculations. In the quark model, the $\chi_{cJ}$ ($J=0,1,2$) mesons are the $^3P_J$ charmonium states. Experimentally and theoretically, their decays are not as extensively studied as those of the vector charmonium
states $J/\psi$ and $\psi(3686)$. At BESIII, the $\chi_{cJ}$ mesons cannot be produced directly through $e^+ e^-$ annihilation processes, but can be accessed via the radiative $\psi(3686)$ decays with a branching fraction of 9\%~\cite{ref::baifen9}, which provides a good opportunity to investigate the properties of these states.

To date, there have been studies of $\chi_{cJ}$ decays involving $p\bar p$ pairs and meson pairs~\cite{ref::pppipi,ref::pppi0pi0,ref::ppkk}, as well as hyperon pairs~\cite{ref::lambdalambdabar,ref::sigmasigmabar} which can be described by the color octet mechanism~\cite{ref::oct1,ref::oct2}.
There is, however, only limited information on decays of $ \chi_{cJ}$ into a $ p \bar p$ pair and three mesons, both in theory and in experiment. Measuring branching fractions of these decays is therefore crucial for a deeper understanding of $\chi_{cJ}$ decay mechanisms and for identifying potential intermediate-state decays. Besides, an enhancement in the $ p \bar p$ invariant mass spectrum has been observed in the study of $J/\psi \to \gamma p\bar p$ ~\cite{ref::ppbar1,ref::ppbar2}. Investigating $ \chi_{cJ}$ decays into $p \bar p$ accompanied by mesons is helpful to explore possible threshold enhancements in different processes.

\setlength{\parskip}{1ex}
In this paper, we present the first experimental studies of $\chi_{cJ}\to p \bar p K^{0}_{S} K^{-} \pi^{+} + c.c.$ by analyzing $(27.12\pm0.14)\times10^8$ $\psi(3686)$ events~\cite{ref::psip-num-inc} collected with the BESIII detector.


\section{BESIII DETECTOR AND MONTE CARLO SIMULATION}
\label{sec:BES}
The BESIII detector~\cite{ref::detector} records symmetric $e^+e^-$ collisions provided by the BEPCII storage ring~\cite{ref::collider}
in the center-of-mass energy range from 2.0 to 4.95~GeV,
with a peak luminosity of $1.1 \times 10^{33}\;\text{cm}^{-2}\text{s}^{-1}$
achieved at $\sqrt{s} = 3.773\;\text{GeV}$.

The cylindrical core of the BESIII detector covers 93\% of the full solid angle and consists of a helium-based
 multilayer drift chamber~(MDC), a plastic scintillator time-of-flight
system~(TOF), and a CsI(Tl) electromagnetic calorimeter~(EMC),
which are all enclosed in a superconducting solenoidal magnet
providing a 1.0~T magnetic field.
The solenoid is supported by an
octagonal flux-return yoke with resistive plate counter muon
identification modules interleaved with steel.
The charged-particle momentum resolution at $1~{\rm GeV}/c$ is
$0.5\%$, and the
${\rm d}E/{\rm d}x$
resolution is $6\%$ for electrons
from Bhabha scattering. The EMC measures photon energies with a
resolution of $2.5\%$ ($5\%$) at $1$~GeV in the barrel (end cap)
region. The time resolution in the TOF barrel region is 68~ps, while
that in the end cap region was 110~ps.
The end-cap TOF system was upgraded in 2015 using multi-gap resistive plate chamber technology, providing a time resolution of 60~ps~\cite{Tof1,Tof2,Tof3}. Around 80\% of the data used in this analysis benefits from this upgrade.

Simulated data samples produced with a {\sc
geant4}-based~\cite{Geant4} Monte Carlo (MC) package, which
includes the geometric description of the BESIII detector and the
detector response, are used to determine detection efficiencies
and to estimate backgrounds. The simulation models the beam
energy spread and initial state radiation (ISR) in the $e^+e^-$
annihilations with the generator {\sc
kkmc}~\cite{Jadach01}.

The inclusive MC sample includes the production of the
$\psi(3686)$ resonance, the ISR production of the $J/\psi$, and
the continuum processes incorporated in {\sc
kkmc}~\cite{Jadach01}.
All particle decays are modelled with {\sc
evtgen}~\cite{Lange01,Lange02} using branching fractions
either taken from the
Particle Data Group~\cite{ref::pdg2022}, when available,
or otherwise estimated with {\sc lundcharm}~\cite{Lundcharm00}.
Final state radiation
from charged final state particles is incorporated using the {\sc
photos} package~\cite{PHOTOS}.
An inclusive MC sample containing $2.7\times10^{9}$ generic $\psi(3686)$ decays is used to study background.
To account for the effect of intermediate
resonance structures on the efficiency, each of the $\chi_{c0,1,2}$
decays is modeled using 1 million mixed signal
MC samples, in which the dominant decay modes contain resonances, such as $\Lambda(1520)$ generated by phase-space model~(PHSP) and $K^*$ generated by a vector meson to a pair of scalar particles model~(VSS), which are mixed with the PHSP signal MC samples. The mixing ratios are determined by examining the corresponding invariant mass spectrum as discussed in Section V.

\section{EVENT SELECTION}
\label{sec:selection}

We reconstruct the events involving the charmonium
transitions $\psi(3686)\to\gamma\chic{J}$, followed by the hadronic
decays $\chic{J}\to p \bar p K^0_S K^- \pi^+ + c.c.$. To identify signal events, we require a minimum of six charged tracks and at least one photon candidate.

Photon candidates are identified using showers in the EMC. The deposited energy of each shower must exceed 25~MeV in the barrel region ($|\cos \theta|< 0.80$, where $\theta$ denotes the
polar angle with respect to the z-axis, the symmetry axis of the MDC) and more than 50~MeV in the end cap region ($0.86 <|\cos \theta|< 0.92$). To exclude showers originating from charged tracks, the angle between the EMC shower and the position of the closest charged track at the EMC must be greater than 10 degrees as measured from the interaction point (IP). To suppress electronic noise and unrelated showers, the difference between the EMC time and the event start time is required to be within [0, 700]\,ns.

The charged tracks except those from $ K^0_S$ detected in the MDC are required to be within a polar angle ($\theta$) range of $|\rm{cos\theta}|<0.93$, $|V_{z}|<10$ cm, and $|V_{xy}|<1$ cm, where $|V_{z}|$ is the distance of closest approach to the IP along the z-axis, and $|V_{xy}|$ is the distance to the
transverse plane.

Particle identification~(PID) for charged tracks combines measurements of the d$E$/d$x$ in the MDC and the flight time in the TOF to form likelihoods $\mathcal{L}(h)$ for each hadron hypothesis $h~(h=p, K, \pi)$.
The charged tracks with $\mathcal{L}(p)>\mathcal{L}(\pi)$ and $\mathcal{L}(p)>\mathcal{L}(K)$ are assigned as protons, remaining tracks with $\mathcal{L}(K)>\mathcal{L}(\pi)$ are identified as kaons, and the rest are identified as pions.

The $ K^0_S$ candidate is reconstructed from two oppositely charged tracks satisfying $|V_{z}|<$ 20 cm, with no constraint applied in the $xy$ plane and PID requirement on charged pions. The first vertex fit is performed on any combination of $\pi^+ \pi^-$ and $\chi^2$ is required to be less than 200. The second vertex fit is applied to further suppress the mis-combination, requiring the decay length divided by its uncertainty to be larger than two .~The $K^{0}_{S}$ signal region is set to be $|M_{\pi^+ \pi^-}-m_{K^{0}_{S}}|<0.012~\mathrm{GeV/}$${c}^2$, where $m_{K^{0}_{S}}$ is the $K^0_S$ nominal mass~\cite{ref::pdg2022}.~The $K^{0}_{S}$ sideband region is set to be $0.020~\mathrm{GeV/}$${c}^2<|M_{\pi^+ \pi^-}-m_{K^{0}_{S}}|<0.044~\mathrm{GeV/}$${c}^2$.

A four-momentum conservation constraint (4C) kinematic fit is applied to the events under the hypothesis of $e^+ e^- \to p \bar p K^0_S K^- \pi^+ \gamma + c.c.$. If more than one combination survives in an event, the one with the smallest $\chi^{2}$ from the 4C fit, $\chi_{\rm 4C}^{2}$ is retained.~The requirement on $\chi^2_{\rm 4C}$ is optimized with the Figure of Merit (FOM) defined as $ \mathrm{FOM} = \mathit{S}/\sqrt{\mathit{S}+\mathit{B}}.$ Here $S$ is the number of events from the signal MC sample, normalized according to the pre-measured branching fractions without $\chi^2$ requirement;
$B$ is the number of background events from the inclusive MC sample, normalized to the data size. From the optimization, we choose $\chi^2_{\rm 4C}<50$ as the nominal requirement.

The potential background components from $\psi(3686)$ decays are studied by analyzing the inclusive MC sample with the generic
event type analysis tool, TopoAna~\cite{topo}. No significant peaking backgrounds with both $\chi_{cJ}$ and $K^0_S$ are observed.
Furthermore, the possible continuum background contribution is
evaluated by examining the data sample taken at $\sqrt s =$ 3.650 GeV, corresponding to an integrated luminosity of 0.4 fb$^{-1}$~\cite{lum}.
Since only a few events survive the selection criteria, the contribution is also ignored.

\section{Branching fraction}

\begin{figure*}[htbp]
	\centering
	\includegraphics[width=8cm]{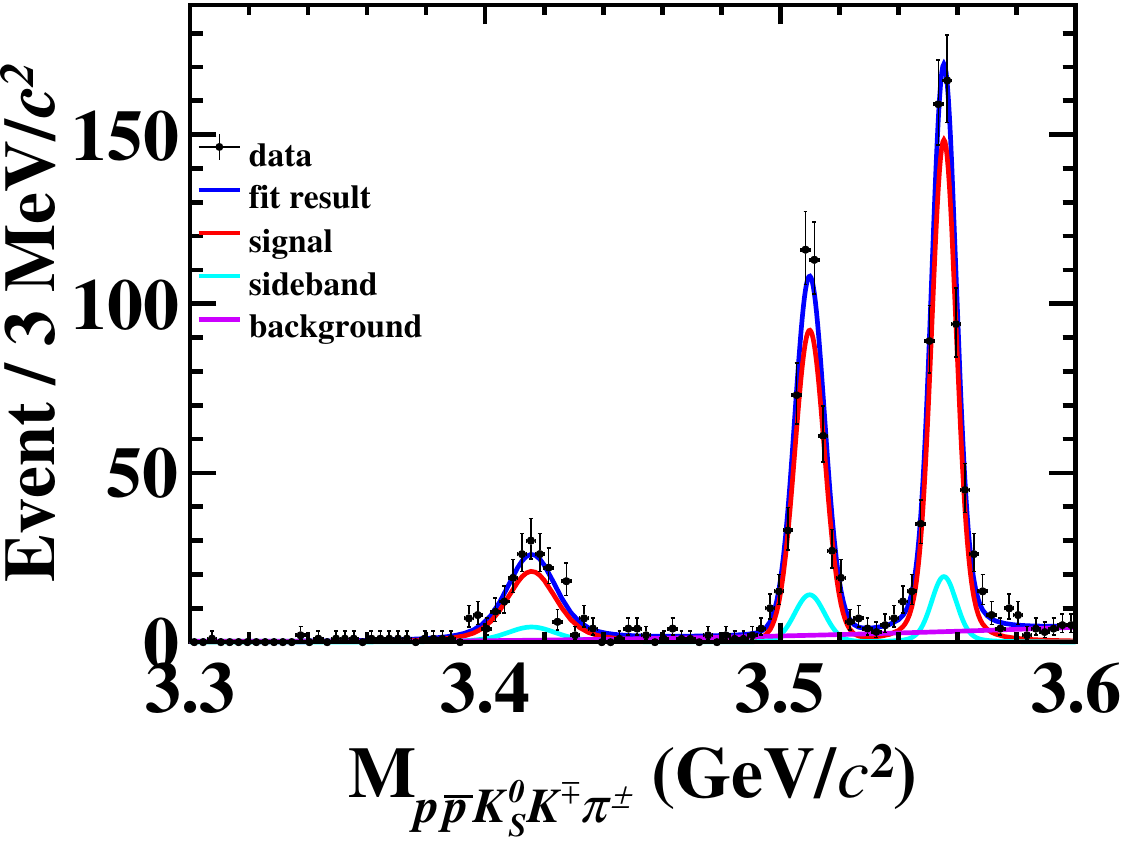}
	\includegraphics[width=8cm]{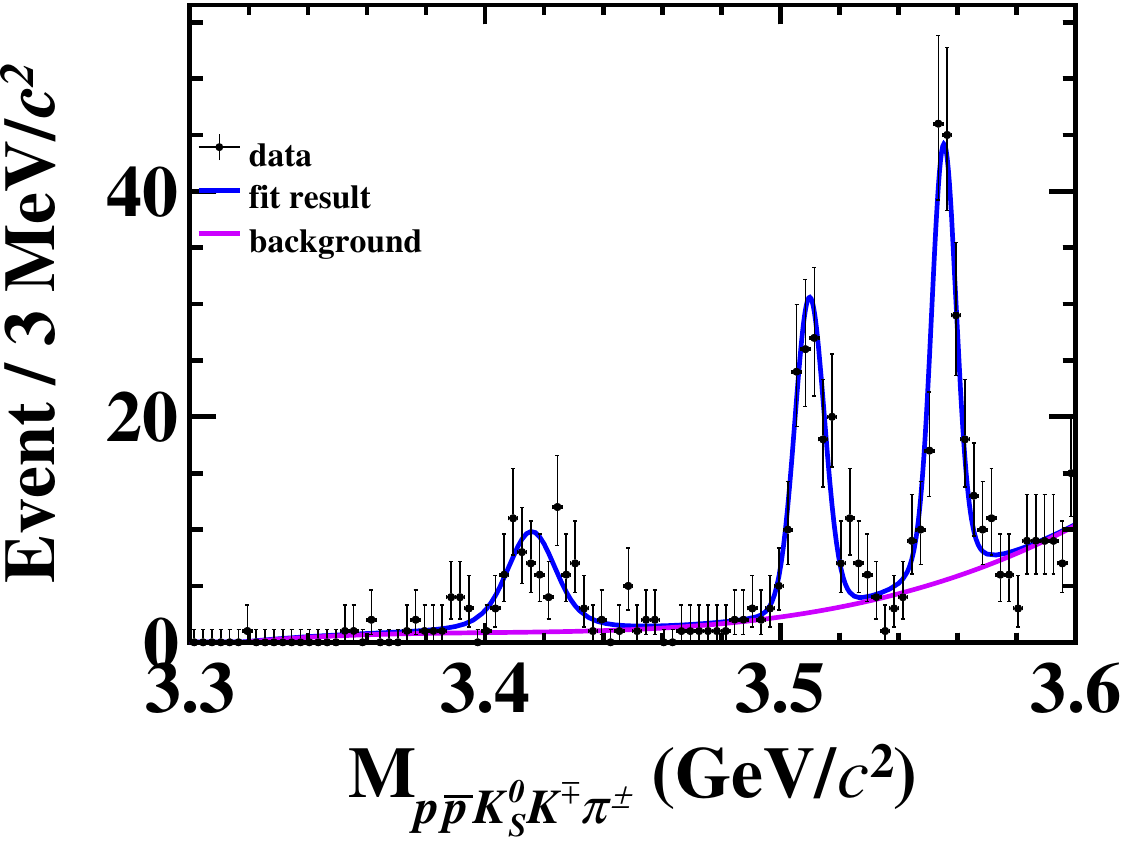}
	\caption{Simultaneous fit to the $M_{p \bar p K^{0}_{S} K^{\pm} \pi^{\mp} }$ distributions of the accepted candidates of $\chi_{cJ}\to p \bar p K^{0}_{S} K^{\pm} \pi^{\mp}$
		in data in the $ K^{0}_{S}$ signal~(left) and sideband~(right) regions (shown as dots with error bars). The blue solid curves are the fit results. The red curve is the signal fit result. The cyan curve is the sideband fit result. The purple curve is the fitted background.}
	\label{tab:k-pi+fit_sig}
\end{figure*}

To determine the signal yields, unbinned maximum likelihood fits are performed simultaneously to the $ M_{p \bar p K^0_S K^\mp \pi^\pm}$ distributions in both the $K^0_S$ signal and sideband regions of the accepted candidates. In the fit, the signal shape of $\chi_{cJ}$ for both the signal and sideband regions is described by a Breit-Wigner function convolved with a Gaussian resolution function with free parameters. The Gaussian resolution functions are independent for three $\chi_{cJ}$ states, but they are shared for the $K^0_S$ signal and sideband regions for each $\chi_{cJ}$ state.  The mass and width of each Breit-Wigner function are
fixed to the corresponding PDG values \cite{ref::pdg2022}. The background shape is described by a polynomial function 
in both the $K^0_S$ signal and the sideband regions. Events in the $K^0_S$ sideband regions are renormalized using a scale factor of 0.5, which is the background ratio between $K^0_S$ signal and sideband regions, to estimate possible non-peaking background without $K^0_S$.~Figure \ref{tab:k-pi+fit_sig} shows the fit result. From this fit, the signal yields of $\chi_{c0}$, $\chi_{c1}$ and $\chi_{c2}$ ($N_{\chi_{cJ}}^{\rm obs}$) are determined to be $172.6\pm17.6$, $395.6\pm23.2$, and $573.8\pm28.3$, respectively. The statistical significances exceed $10\sigma$ for each $\chi_{cJ}$ decay, as determined by $\Delta(\rm{ln}\mathcal{L}) = \rm{ln}\mathcal{L}_{\rm max} - \rm{ln}\mathcal{L}_{0}$ and $\Delta ndf = 1$.~Here, $\mathcal{L}_{\rm max}$ and $\mathcal{L}_{0}$ are the maximum likelihoods with and without
the signal component in the fit; $\Delta ndf$ is the variation of number of degrees of freedom.

\begin{figure*}[htbp]
	\centering
	\includegraphics[width=14cm]{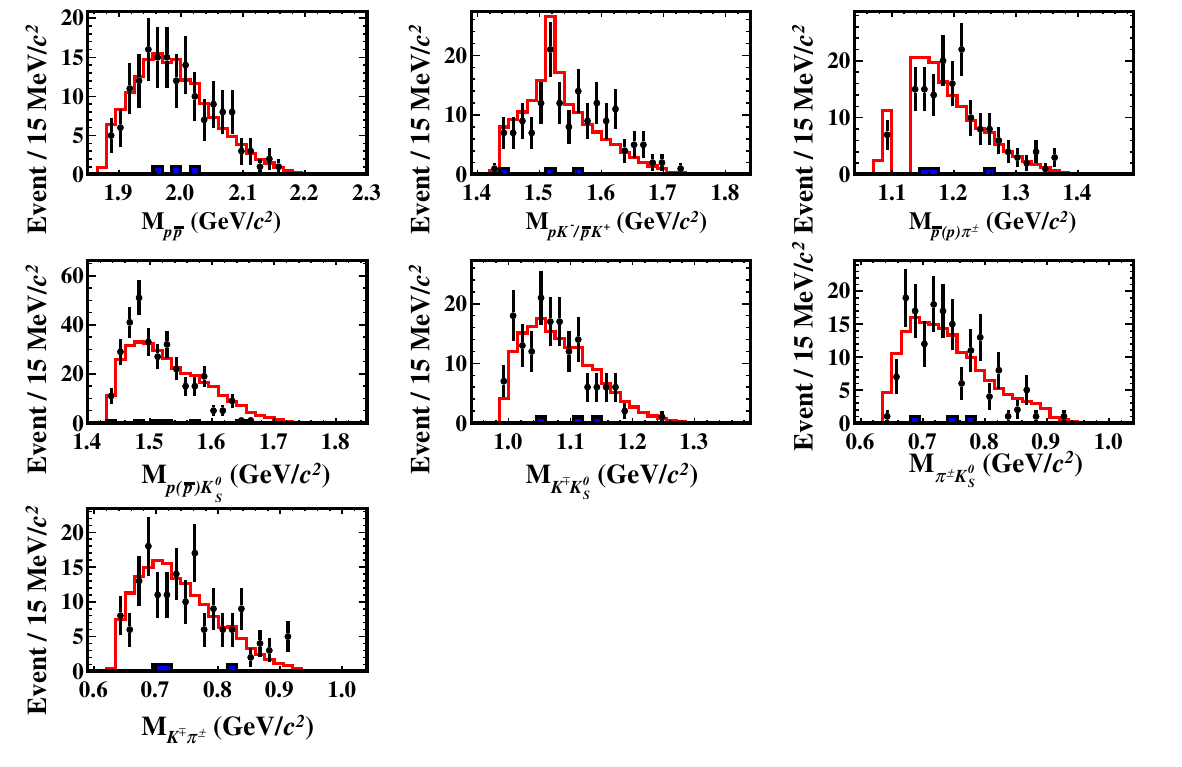}	
	\caption{Comparisons of the invariant masses of two-body combinations of the accepted candidates for $\chi_{c0}\to p\bar p K^0_S K^\mp \pi^\pm$ in data and mixed signal MC sample. The black dots with error bars are data. The white histograms with red lines are mixed signal MC samples. The blue histograms are background coming form inclusive MC sample, in the $\chi_{c0}$ mass region with $M_{p\bar p K^0_SK^\mp\pi^\pm} \in[3.394, 3.434]$ GeV/$c^2$.}
	\label{tab:con0}
\end{figure*}

\begin{figure*}[htbp]
	\centering
	\includegraphics[width=14cm]{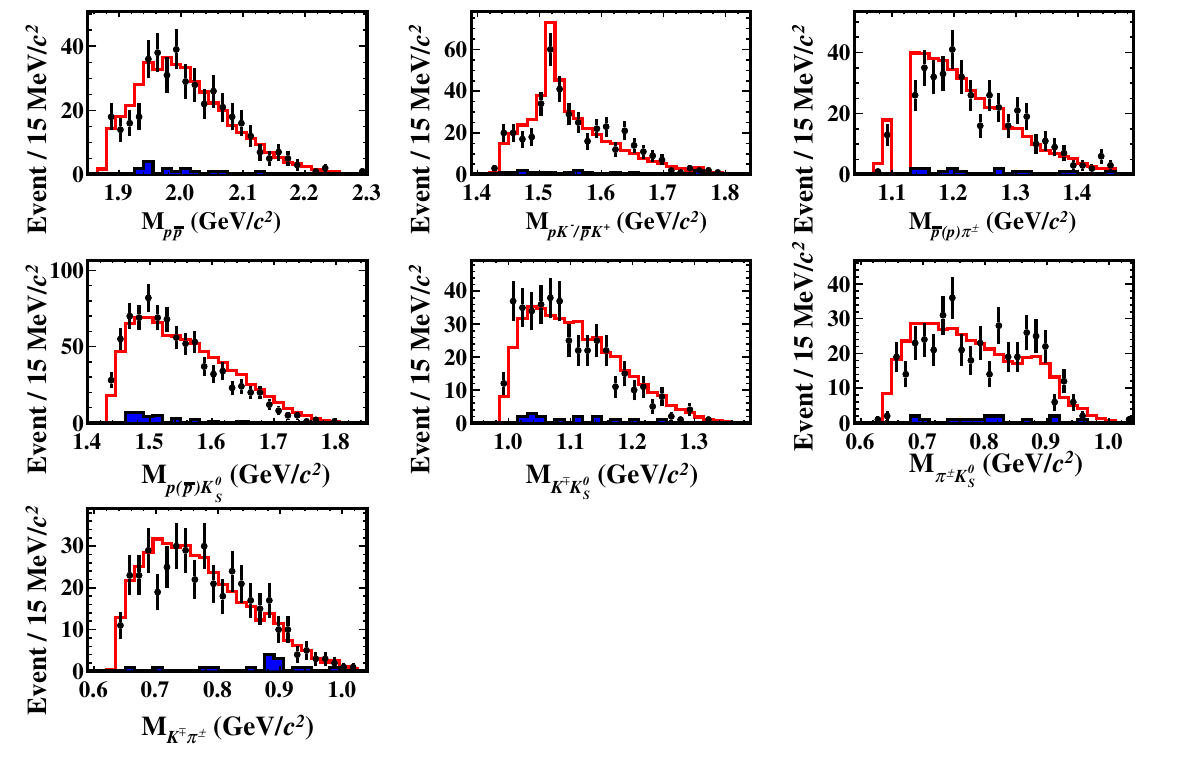}
	\caption{Comparisons of the invariant masses of two-body combinations of the accepted candidates for $\chi_{c1}\to p\bar p K^0_S K^\mp \pi^\pm$ in data and mixed signal MC sample. The black dots with error bars are data. The white histograms with red lines are mixed signal MC samples. The blue histograms are background coming form inclusive MC sample, in the $\chi_{c1}$ mass region with $M_{p\bar p K^0_SK^\mp\pi^\pm} \in[3.496, 3.526]$ GeV/$c^2$.}
	\label{tab:con1}
\end{figure*}

\begin{figure*}[htbp]
	\centering

	\includegraphics[width=14cm]{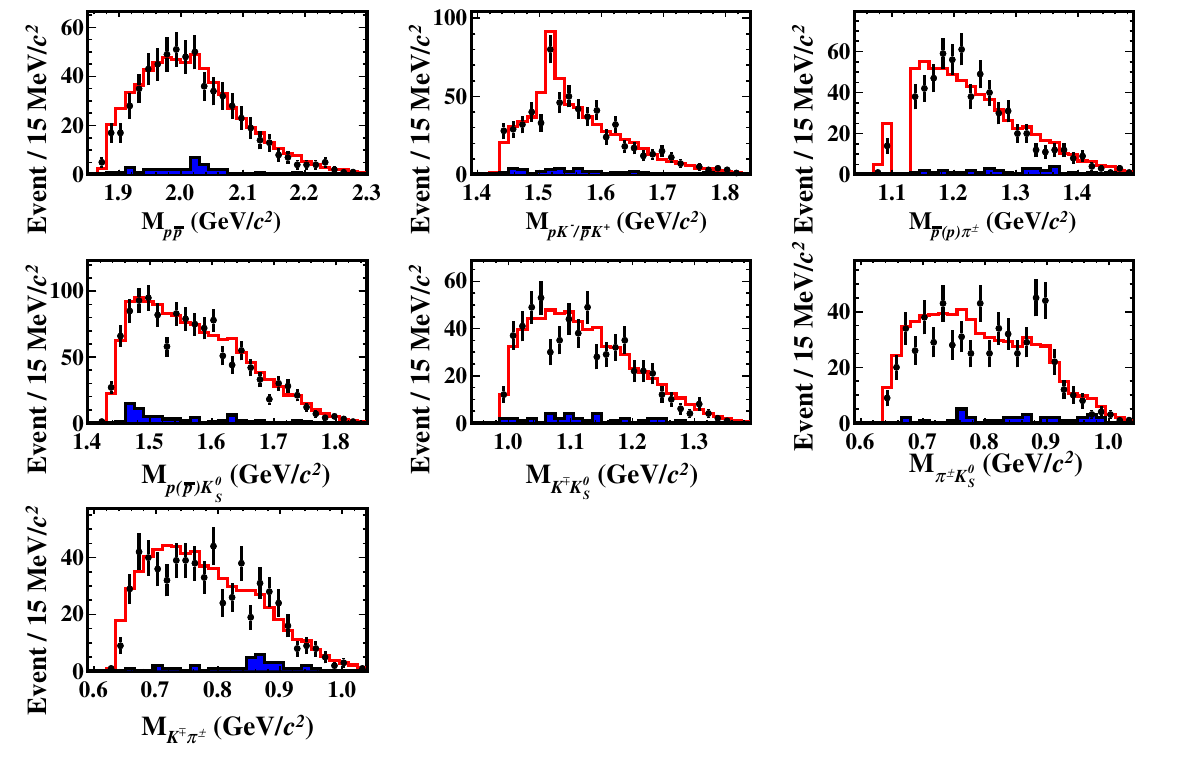}
	\caption{Comparisons of the invariant masses of two-body combinations of the accepted candidates for $\chi_{c2}\to p\bar p K^0_S K^\mp \pi^\pm$ in data and mixed signal MC sample. The black dots with error bars are data. The white histograms with red lines are mixed signal MC samples. The blue histograms are background coming form inclusive MC sample, in the $\chi_{c2}$ mass region with $M_{p\bar p K^0_SK^\mp\pi^\pm} \in[3.541, 3.571]$ GeV/$c^2$.}
	\label{tab:con2}
\end{figure*}

\begin{figure*}[htbp]
	\centering
	\includegraphics[width=8cm]{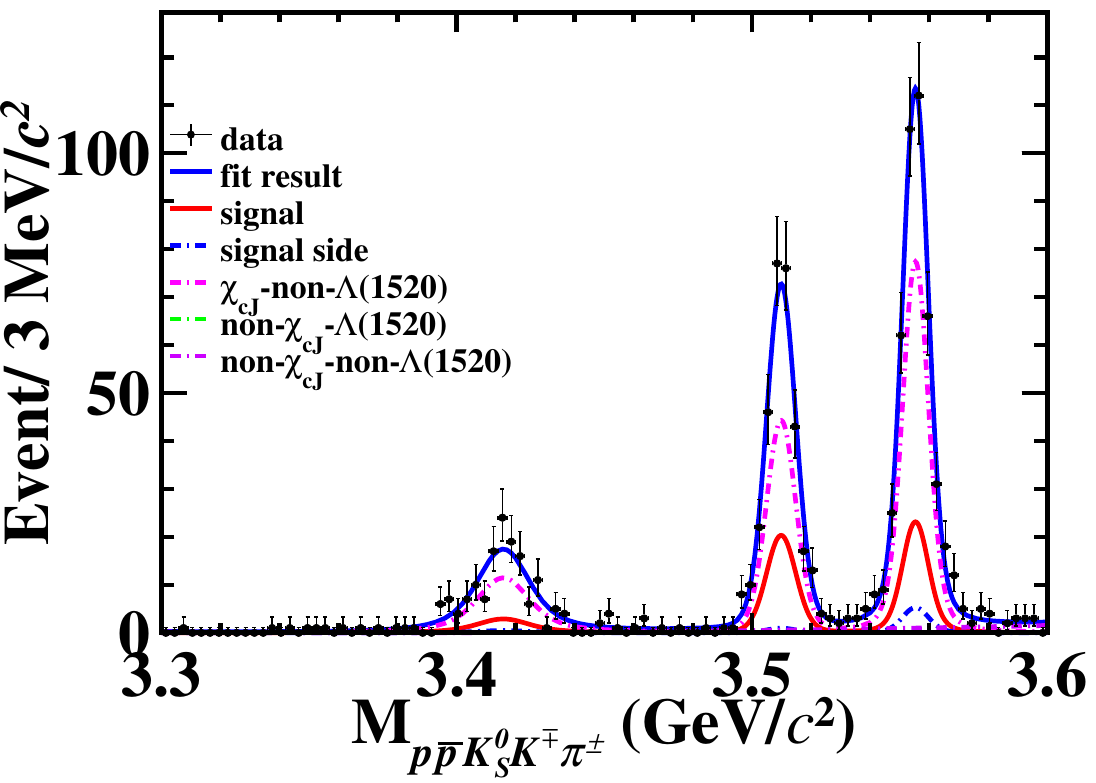}
	\includegraphics[width=8cm]{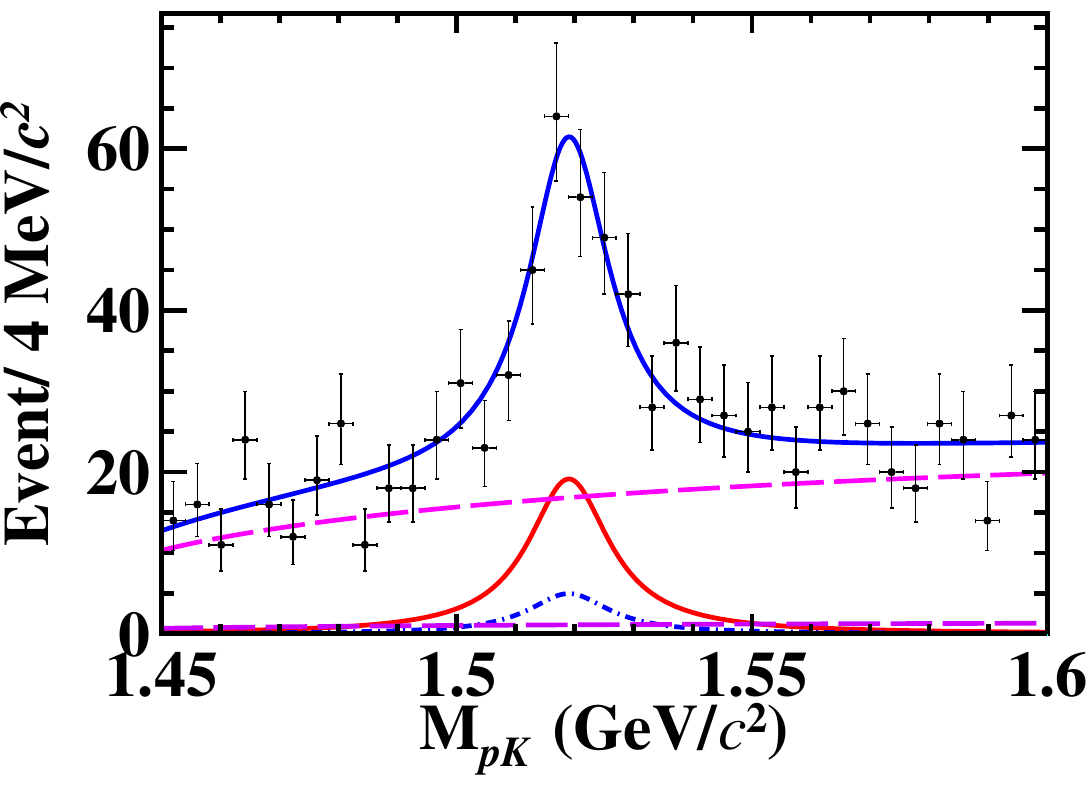}
	\includegraphics[width=8cm]{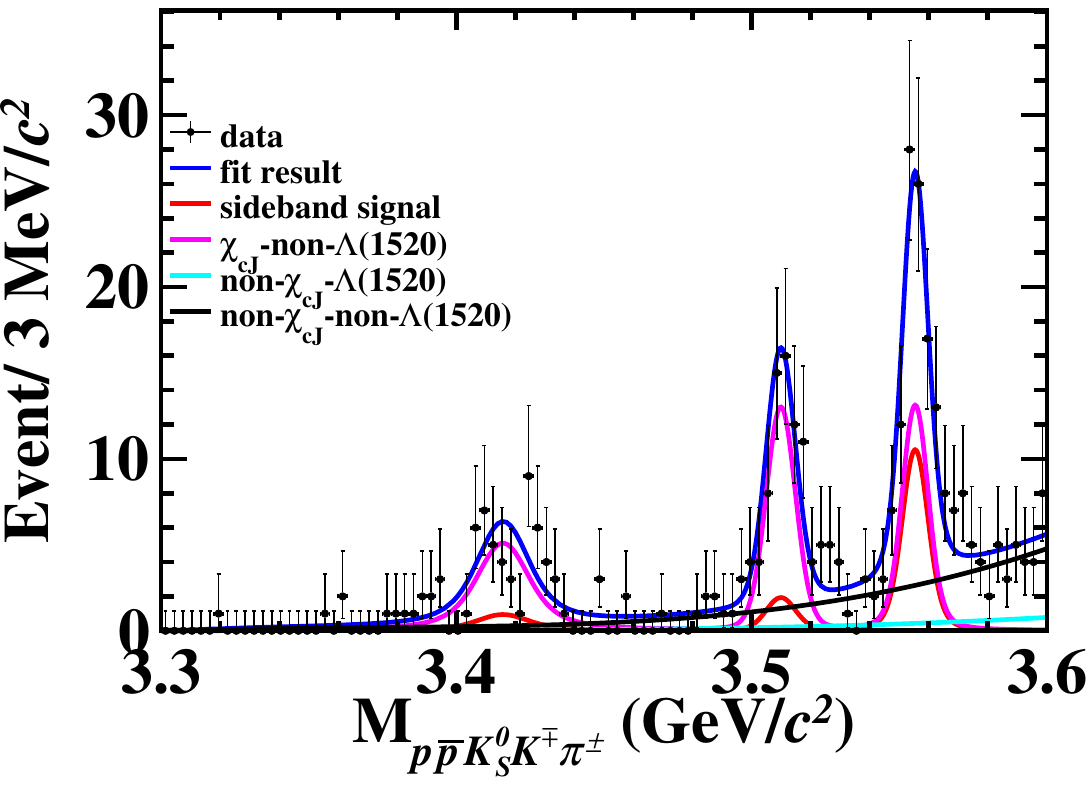}
	\includegraphics[width=8cm]{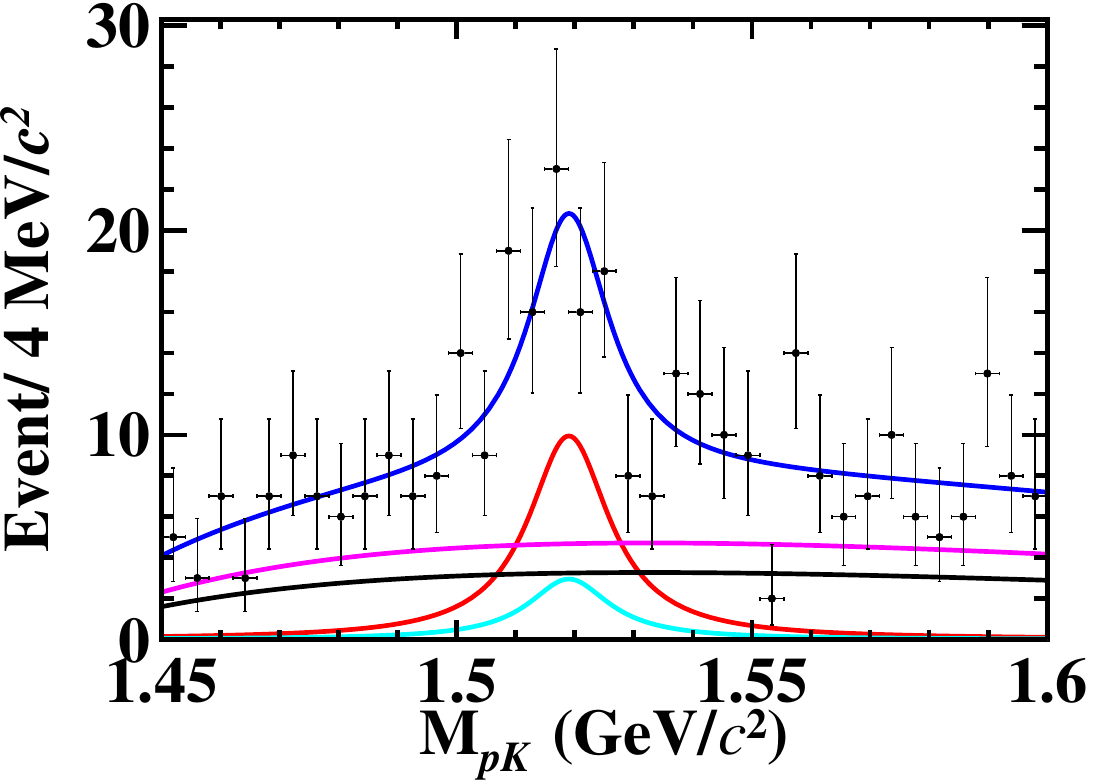}
	\caption{The 2D simultaneous fit to the $M_{p \bar p K^{0}_{S} K^{\mp} \pi^{\pm} }$ and $ M_{ p K} $ distributions in data of the $K^{0}_{S}$ signal~(top) and sideband~(bottom) regions (shown as dots with error bars). For the top figures,~the blue solid curves are the fit results.~The red solid curves are the fitted signal shapes. The blue dashed curves are the normalized sideband contributions. The pink dashed curves are $\chi_{cJ}-$non-$\Lambda(1520)$ events. The cyan dashed curves are non-$\chi_{cJ}-\Lambda(1520)$ events.~The purple dashed curves are  non-$\chi_{cJ}-$non-$\Lambda(1520)$ events.~For the bottom figures, the blue solid curves are the fit results. The red solid curves are the fitted signal shapes. The pink dashed curves are $\chi_{cJ}-$non-$\Lambda(1520)$ events.~The cyan dashed curves are non-$\chi_{cJ}-\Lambda(1520)$ events.~The purple dashed curves are non-$\chi_{cJ}-$non-$\Lambda(1520)$ events.}
	\label{tab:2Dk-pi+sig}
\end{figure*}

Considering the $p K$ invariant mass spectrum in Figs. \ref{tab:con0}, \ref{tab:con1}, and \ref{tab:con2}, one can observe resonance sub-processes $\chi_{cJ}\to \bar{p} \Lambda(1520) K^0_S \pi^{+} + c.c.$ in the decays $\chi_{cJ} \to p \bar{p} K^0_S K^- \pi^+ + c.c.$. The signal yields of $\chi_{cJ}\to \bar{p} \Lambda(1520) K^0_S \pi^{+} + c.c.$ are determined from a two-dimensional~(2D) simultaneous likelihood fit to the distributions of
$M_{pK}$(including $M_{pK^{-}}$ and $M_{\bar{p}K^{+}}$) versus $M_{p \bar{p}K^{0}_{S}K^{\pm}\pi^{\mp}}$ of events in the $K^0_S$ signal and sideband regions of the accepted candidates in data. 
The signal shape of $\Lambda(1520)$ is described by a Breit-Wigner function convolved with a Gaussian resolution function with free parameters. The Gaussian resolution functions are independent for three $\chi_{cJ}$ states, but they are shared for the $K^0_S$ signal and sideband regions for each $\chi_{cJ}$ state. The mass and width of each Breit-Wigner function are fixed to the corresponding PDG values~\cite{ref::pdg2022}.
The combinatorial background in the $M_{pK}$ distributions is described by an ARGUS function. 
The shapes of the $\chi_{cJ}$ signals and the combinatorial backgrounds in the $M_{p \bar{p} K^0_SK\pi}$ distributions follow the same descriptions used in the one-dimensional simultaneous likelihood fit.~Events in the $K^0_S$ sideband regions are normalized with a scale factor, which is 0.5 for three $\chi_{cJ}$ states and determined by the ratio of $K^0_S$ signal region to the sideband region, to estimate potential peaking background without $K^0_S$.~Figure \ref{tab:2Dk-pi+sig} shows the 2D fit results in the signal and sideband regions. From this fit, the signal yields of $\chi_{c0,1,2}\to  \bar p \Lambda(1520) K^{0}_{S} \pi^+ +c.c.$, $N_{\chi_{cJ}}^{\rm obs}$, are determined to be $27.0^{+11.4}_{-10.7}$, $88.2^{+17.3}_{-16.5}$, and $93.8^{+20.1}_{-19.3}$, respectively,
with statistical significances of $3.3\sigma$, $5.7\sigma$, and $7.0\sigma$, respectively. Besides, no significant enhancement around the $p\bar{p}$ mass threshold is observed, as shown in Figs. \ref{tab:con0}, \ref{tab:con1} and \ref{tab:con2}.

The efficiencies of detecting $\psi(3686)\to\gamma\chi_{cJ}$ with $\chi_{cJ}\to p \bar p K^0_S K^{\mp} \pi^{\pm}$
are determined using mixed signal MC samples with fractions of 
$\chi_{cJ}\to  p \bar p K^0_S K^{-} \pi^{+} + c.c.$, 
$\chi_{cJ}\to \bar{p} \Lambda(1520) K^0_S \pi^{+} + c.c.$, and 
$\chi_{cJ}\to p \bar p K^{*+} K^{-} + c.c.$, as summarized in Table \ref{tab:weight}. The fractions of $\chi_{cJ}\to \bar p  \Lambda(1520) K^{0}_{S} \pi^{+}+c.c.$ are determined by a 2D simultaneous fit shown in Fig.~\ref{tab:2Dk-pi+sig}. The fractions of $\chi_{cJ}\to p \bar p K^{*\pm} K^{\mp}$ are estimated from a preliminary 2D fit to the $K^{0}_{S}\pi^{\pm}$ invariant mass spectra. However, due to its low significance, we have chosen not to report it. Figures \ref{tab:con0},~\ref{tab:con1}, and \ref{tab:con2} show a comparison of the invariant masses of two-body combinations of the accepted candidates for $\chi_{cJ}\to p\bar p K^0_S K^- \pi^+ + c.c.$ in data and mixed signal MC sample. The branching fractions of $\chi_{cJ}\to p\bar p K^{-} K^{*+} + c.c.$ are not reported 
due to significant background contamination, and the $\chi_{cJ}\to p\bar p K^0_S \bar{K^{*0}}(K^{*0})$ decays are not included in mixed signal MC samples due to the absence of a significant signal.~The obtained detection efficiencies for $\chi_{cJ}\to p \bar p K^0_S  K^\mp \pi^\pm$ and $\chi_{cJ}\to \bar{p} \Lambda(1520) K^0_S \pi^{+} + c.c.$
are shown in Tables \ref{tab:Branchingk-pi+} and \ref{tab:Branchinglambda1520}, respectively.

\begin{table}[htbp]
	\centering	
	\caption{The fractions (in \%) of different subprocesses in $\chi_{cJ}\to p \bar p K^{0}_{S} K^{\mp} \pi^{\pm}$
		, where uncertainties are statistical only.}
	\begin{tabular}{lccc}

		\hline
		$ \chi_{cJ}$ & $ p \bar p K^{0}_{S} K^{-} \pi^{+} + c.c.$ & $ \bar p \Lambda(1520) K^{0}_{S} \pi^+ +c.c.$ & $ p \bar p K^{-} K^{*+} + c.c.$ \\
		\hline
		
		$ \chi_{c0}$ & $72.4^{+7.6}_{-7.1}$ & $15.6^{+6.6}_{-6.2} $ & $12.0^{+3.7}_{-3.5} $ \\

		$ \chi_{c1}$ & $62.7^{+5.9}_{-5.7}$ & $22.3^{+4.4}_{-4.2} $& $15.0^{+4.0}_{-3.9} $ \\

		$ \chi_{c2}$ & $73.7^{+4.8}_{-4.7}$ & $16.3^{+3.5}_{-3.4} $ &$10.0^{+3.3}_{-3.2} $\\
	
		\hline
		
	\end{tabular}
	
	\label{tab:weight}
\end{table}

\label{sec:mc}

The product of branching fractions of $\psi(3686)\to \gamma\chi_{cJ}$ and $\chi_{cJ}\to f$ is calculated as
\begin{equation}
\begin{aligned}
{\mathcal B}(\psi(3686)\to \gamma\chi_{cJ}) \cdot \mathcal{B}(\chi_{cJ}\to f) 
=\frac{N^{\rm obs}_{\chi_{cJ}}}{N_{\psi(3686)}\cdot\epsilon \cdot \mathcal B( sub)},
\end{aligned}
\end{equation}	
where $\mathcal{B}(\chi_{cJ}\to f)$ is the branching fractions of $\chi_{cJ}$ decaying to the final state, $N_{\psi(3686)}$ is the total number of $\psi(3686)$ events in data, and $\epsilon$ is the detection efficiency. The variable $\mathcal B( sub)$ is $\mathcal B( K^0_{S} \to \pi^+ \pi^-)$ for the calculation of the branching fractions of $\chi_{cJ} \to p \bar{p} K^0_S K^{-} \pi^{+} + c.c.$, and $\mathcal B( sub)$ is $\mathcal B( K^0_{S} \to \pi^+ \pi^-) \cdot \mathcal B( \Lambda(1520) \to p K^-)$ for the calculation of the branching fractions of $\chi_{cJ} \to \bar{p} \Lambda(1520) K^0_S \pi^+ + c.c$ .
~Combining the branching fractions of $\psi(3686)\to\gamma\chi_{cJ}$ quoted from the PDG~\cite{ref::pdg2022}, the branching fractions of the $\chi_{cJ} \to f$ decays are determined. All the obtained results for $\chi_{cJ}\to p\bar p K^0_S K^- \pi^+ + c.c.$ and $\chi_{cJ}\to p\Lambda(1520)  K^0_S \pi^++c.c.$ are summarized in Tables \ref{tab:Branchingk-pi+} and  \ref{tab:Branchinglambda1520}, respectively.

\begin{table*}[htbp]\small
	\caption{The quantities used for calculating the branching fractions of $\chi_{cJ} \to p \bar p K^{0}_{S} K^{\mp} \pi^{\pm}$. The first uncertainties are statistical and the second systematic.}
	\centering
		\begin{tabular}{cccc}
			\hline
			& $ \chi_{c0}$ & $ \chi_{c1}$ & $ \chi_{c2} $ \\ \hline
		$N_{\chi_{cJ}}^{\rm obs}$ & $172.6\pm17.6$ & $395.6\pm23.2$ &$573.8\pm28.3$ \\  
		$\epsilon~(\%)$    & $3.60\pm0.02$ & $5.20\pm0.02$ & $5.70\pm0.02$ \\ 
			$\mathcal{B}(\psi(3686)\to\gamma\chi_{cJ})\cdot \mathcal{B}(\chi_{cJ}\to p \bar p K^0_S K^- \pi^+ + c.c.)~(\times 10^{-6})$ & $2.55\pm0.26\pm0.31$& $4.05\pm0.24\pm0.39$& $5.36\pm0.26\pm0.43$ \\
			
			$\mathcal{B}(\chi_{cJ}\to p \bar p K^0_S K^- \pi^+ + c.c.)~(\times 10^{-5})$ & $2.61\pm0.27\pm0.32$& $4.16\pm0.24\pm0.46$& $5.63\pm0.28\pm0.46$ \\   
			\hline
		\end{tabular}
	\label{tab:Branchingk-pi+}
\end{table*}

\begin{table*}[htbp]
	\caption{The quantities used for calculating the branching fractions of $\chi_{cJ}\to\bar p \Lambda(1520) K^{0}_{S} \pi^{+}+c.c.$.
The first uncertainties are statistical and the second systematic.}
	\centering
		\begin{tabular}{cccc}
			
			\hline
			& $ \chi_{c0}$ & $ \chi_{c1}$ & $ \chi_{c2} $ \\ \hline
			$N_{\chi_{cJ}}^{\rm obs}$ & $27.0^{+11.4}_{-10.7}$ & $88.2^{+17.3}_{-16.5}$ &$93.8^{+20.1}_{-19.3}$  \\  
			$\epsilon~(\%)$    & $4.06\pm0.06$ & $5.27\pm0.07$ & $5.71\pm0.07$\\
			$\mathcal{B}(\psi(3686)\to\gamma\chi_{cJ})\cdot \mathcal{B}(\chi_{cJ}\to \bar{p} \Lambda(1520) K^0_S \pi^{+} + c.c.)~(\times 10^{-6})$ & $1.57^{+0.66}_{-0.62}\pm0.22$ & $3.96^{+0.77}_{-0.74}\pm0.50$ & $3.89^{+0.83}_{-0.80}\pm0.39$ \\   
			$\mathcal{B}(\chi_{cJ}\to \bar{p} \Lambda(1520) K^0_S \pi^{+} + c.c.)~(\times 10^{-5})$ & $1.61^{+0.68}_{-0.64}\pm0.23$ & $4.06^{+0.80}_{-0.76}\pm0.52$ & $4.09^{+0.87}_{-0.84}\pm0.42$ \\  
			\hline
		\end{tabular}
	\label{tab:Branchinglambda1520}
\end{table*}

\section{SYSTEMATIC UNCERTAINTY}
\label{sec:systematics}

The systematic uncertainties in the branching fraction measurements arise from various sources, as summarized in Table~\ref{tab:systematics}. These uncertainties are estimated and discussed below.
\begin{table*}[htbp]
	\centering
  	\caption{Relative systematic uncertainties (in \%) in the branching fraction measurements.
  	The numbers before and after ``/" are the uncertainties for $\chi_{cJ}\to p \bar p K^{0}_{S} K^{-} \pi^{+} + c.c.$ and $\chi_{cJ}\to\bar p \Lambda(1520) K^{0}_{S} \pi^{+}+c.c.$, respectively.}
  \begin{tabular}{lccc}
  
  	\hline
  	Source & $\chi_{c0}$ & $\chi_{c1}$ & $\chi_{c2}$  \\
  	\hline
  	$N_{\psi(3686)}$ & $ 0.5$ & $ 0.5$ & $ 0.5$ \\
  
  	Tracking & $ 4.0$ & $ 4.0$ & $ 4.0$ \\
  	
  	PID      & $ 4.0$ & $ 4.0$ & $ 4.0$ \\
  
  	$K_{S}^{0}$ reconstruction & $ 1.5$ & $ 1.5$ & $ 1.5$\\
  
  	$\gamma$ selection & $ 1.0$ & $ 1.0$ & $ 1.0$\\
  
  	4C kinematic fit & $ 1.0/1.6$ & $ 1.2/0.9$ & $ 2.5/1.5$ \\
  
  	Invariant mass fit & $ 5.8/1.1$ & $ 3.0/2.5$ & $ 1.3/2.3$ \\
  
  	Low end of fit region & $ 4.9/2.6$ & $ 1.5/0.2$ & $ 0.8/0.1$ \\
  	
  	Fit bias & $ 3.0/10.0$ & $ 3.0/10.0$ & $ 3.0/7.0$ \\
  	
  	$\mathcal B(K_{S}^{0} \to \pi^{+} \pi^{-})$ & $ 0.1$ & $ 0.1$ & $ 0.1$ \\
  
  	${\mathcal B}(\Lambda(1520)\to pK^-)$
  	 & $ .../2.2$ & $ .../2.2$ & $ .../2.2$ \\
  
  	MC statistics & $ 0.5/1.4$ & $ 0.4/1.3$ & $ 0.4/1.2$\\
  
  	MC generator & $ 6.8/6.8$ & $ 6.1/2.7$ & $ 3.2/2.1$ \\
  	
  	Imperfect simulation for $M_{pK^0_S}$ & $ 0.5/...$ & $ .../...$ & $ .../...$ \\
  
  	Total for $\mathcal{B}(\psi(3686)\to\gamma\chi_{cJ}) \cdot \mathcal{B}(\chi_{cJ}\to f) $ & $ 12.2/14.1$ & $ 9.7/12.5$ & $ 8.0/10.0$ \\
  
  	Total for $\mathcal{B}(\chi_{cJ} \to f$) & $ 12.4/14.2$ & $ 10.0/12.7$ & $ 8.2/10.2$ \\
  
  	\hline
  
  \end{tabular}
  \label{tab:systematics}
\end{table*}

The total number of $\psi(3686)$ events in data is measured to be $N_{\psi(3686)}=(27.12\pm0.14)\times10^8$ with the inclusive
hadronic data sample as described in Ref.~\cite{ref::psip-num-inc}. The uncertainty of $N_{\rm \psi(3686)}$ is 0.5\%.

The systematic uncertainties of the tracking or PID efficiencies for $K^{\pm},\pi^{\pm}$ and $p(\bar{p})$ are estimated with the control samples of $J/\psi\to K^{*}\bar{K}$ and $J/\psi \to p\bar p \pi^+\pi^-$. They are assigned as 1.0\% for tracking or PID efficiencies per track~\cite{ref::tracking}.

The systematic uncertainty of the $K^0_S$ reconstruction, including tracking efficiency, $K^0_S$ mass window, vertex fit, and second vertex fit, is estimated using the control sample of $J/\psi \rightarrow K^{*\pm} ¯\bar{K}^\mp$ and $J/\psi \rightarrow \phi K^{*\pm} ¯\bar{K}^\mp$. The systematic uncertainty of the $K^0_S$ reconstruction is assigned to be 1.5\%~\cite{ref::kso}.

The systematic uncertainty in the photon detection is assigned as  1.0\% per photon based on the control sample of $J/\psi\to\pi^+\pi^-\pi^0$~\cite{ref::gamma-recon}.

To estimate the systematic uncertainties associated with variations in the sub-resonant fractions of the MC model for the $\chi_{cJ}\to p \bar{p} K^0_S K^{\mp} \pi^{\pm}$ decays, we compare our nominal signal efficiencies with those determined from the mixed signal MC events after varying the relative fractions of the sub-resonant decays, such as $\chi_{cJ}\to p \bar p K^{0}_{S} K^{-} \pi^{+} + c.c.$, $\chi_{cJ}\to \bar{p} \Lambda(1520) K^0_S \pi^{+} + c.c.$, and $\chi_{cJ}\to p \bar p K^{*+} K^{-} + c.c.$, by $\pm 1$ standard deviation, and mixing the $\chi_{cJ}\to p \bar p K^{*0} K^{0}_{S} + c.c.$ component with the same fraction as $\chi_{cJ}\to p\bar p K^{*\pm} K^\mp$. The changes of the signal efficiencies, 6.8\%, 6.1\%, and 3.2\%, are taken as the systematic uncertainties for \chic{0}, \chic{1}, and \chic{2} decays, respectively. The systematic uncertainties of the MC model for $\chi_{c0,1,2}\to \bar{p} \Lambda(1520) K^0_S \pi^{+} + c.c.$ are estimated by mixing the $\chi_{c0,1,2}\to \bar p \Lambda(1520) K^{*+}$ components in individual nominal signal MC samples with fractions by referring to that of $\chi_{cJ}\to p \bar p K^{*+} K^{-} + c.c.$ in the studies of $\chi_{cJ}\to p\bar p K^0_SK^\pm\pi^\mp$. The relative changes in the signal efficiencies, 6.8\%, 2.7\%, and 2.1\%, are assigned as the corresponding systematic uncertainties for \chic{0}, \chic{1}, and \chic{2} decays, respectively.

The systematic uncertainty due to imperfect simulation of $M_{p(\bar p)K^0_S}$ for $\chi_{c0}\to p \bar p K^{0}_{S} K^{-} \pi^{+} + c.c.$ is estimated by mixing a component of $\chi_{c0}\to {\mathcal R} (\to  p K^0_S) \bar p K^-\pi^++c.c.$  in the nominal signal MC sample, in which the mass and width of $\mathcal R$ are set to be 1.485~GeV/$c^2$ and 20 MeV, respectively.~The change in the signal efficiency, 0.5\%, is assigned as the systematic uncertainty. This systematic uncertainty is negligible for other decays. 

The systematic uncertainties of the fit to the $M_{p \bar p K^{0}_{S} K^{\mp} \pi^{\pm}}$ distribution are considered in two parts, the signal and the background shape.
The uncertainties due to the background shape are estimated by varying the order of the polynomial function employed to describe the background. In the case of the 2D simultaneous fit, the background shape in $M_{pK}$ is modified by adjusting the polynomial function.
The uncertainties from the signal shape are estimated by using the MC simulated shapes of $M_{p \bar p K^{0}_{S} K^{\mp} \pi^{\pm}}$ and $M_{pK}$, both of which are convolved with Gaussian functions. The parameters of the smeared Gaussian function are free in the 1D simultaneous fit,
while the parameters of the smeared Gaussian function are fixed at those obtained from the 1D fits for the 2D simultaneous fit.
By combining these two effects in quadrature, the total systematic uncertainties are calculated to be 5.8\%, 3.0\%, and 1.3\%
for $\chi_{c0,1,2}\to p \bar p K^{0}_{S} K^{-} \pi^{+} + c.c.$;
and 1.1\%, 2.5\%, and 2.3\%
for $\chi_{c0,1,2}\to \bar{p} \Lambda(1520) K^0_S \pi^{+} + c.c.$,~respectively.

The systematic uncertainties due to the fit bias are estimated by performing input-output checks based on the inclusive MC sample.
The differences between the measured and input branching fractions are taken as the systematic uncertainties, 
which are 3.0\%,~3.0\%, and 3.0\% for  $\chi_{c0,1,2}\to p \bar p K^{0}_{S} K^{-} \pi^{+} + c.c.$; and 
10.0\%, 10.0\%, and 7.0\% for $\chi_{c0,1,2}\to \bar{p} \Lambda(1520) K^0_S \pi^{+} + c.c.$,~respectively. 

The systematic uncertainties from the low end of the fit range are estimated employing the alternative fit ranges of 
$[3.25,3.60]$ and $[3.35,3.60]$ GeV$/c^{2}$.
The larger differences of branching fractions changed with the fitted signal yields are taken as the systematic uncertainties, which are 
4.9\%, 1.5\% and 0.8\% for $\chi_{c0,1,2}\to p \bar p K^{0}_{S} K^{-} \pi^{+} + c.c.$;
and 2.6\%, 0.2\%, and 0.1\% for $\chi_{c0,1,2}\to \bar{p} \Lambda(1520) K^0_S \pi^{+} + c.c.$,~respectively.

The systematic uncertainty from the 4C kinematic fit is estimated by comparing the signal efficiencies before and after applying a helix parameter correction.
The correction factors are obtained from Ref.~\cite{ref::helixp}.
The changes in the signal efficiencies are assigned as the systematic uncertainties, which are 1.0\%, 1.2\%, and 2.5\% for $\chi_{c0,1,2}\to p \bar p K^{0}_{S} K^{-} \pi^{+} + c.c.$; and 1.6\%, 0.9\%, and 1.5\% for $\chi_{c0,1,2}\to \bar{p} \Lambda(1520) K^0_S \pi^{+} + c.c.$,~respectively.

The systematic uncertainties due to the limited statistics of the signal MC samples
are 0.5\%, 0.4\%, and 0.4\% for  $\chi_{c0,1,2}\to p \bar p K^{0}_{S} K^{-} \pi^{+} + c.c.$,
and 1.4\%, 1.3\%, and 1.2\% for $\chi_{c0,1,2}\to \bar{p} \Lambda(1520) K^0_S \pi^{+} + c.c.$,~respectively.

The uncertainties from the branching fractions quoted from the PDG~\cite{ref::pdg2022} are 2.0\%, 2.5\%, and 2.1\% for $\psi(3686)\to \gamma\chi_{c0,1,2}$,
0.07\% for $\mathcal{B}(K^{0}_{S} \to \pi^+ \pi^-)$, and
2.2\% for $\mathcal{B}(\Lambda(1520) \to p K^-)$.

For each signal decay, the total systematic uncertainty is obtained by adding all individual systematic uncertainties in quadrature.
 
\section{Summary}
By analyzing $(27.12\pm0.14)\times10^8$ $\psi(3686)$ events collected with the BESIII detector operating at the BEPCII collider, 
the decays of $\chi_{c0,1,2} \to p \bar{p} K^0_S K^- \pi^+ + c.c.$ are observed for the first time with statistical significances greater than $10\sigma$.
Their decay branching fractions are determined to be
\begin{equation}
\mathcal{B}(\chi_{c0}\to p \bar p K^{0}_{S} K^{-} \pi^{+} + c.c.)=(2.61\pm0.27\pm0.32)\times10^{-5},
\nonumber
\end{equation}
\begin{equation}
\mathcal{B}(\chi_{c1}\to p \bar p K^{0}_{S} K^{-} \pi^{+} + c.c.)=(4.16\pm0.24\pm0.46)\times10^{-5},
\nonumber
\end{equation} 
\begin{equation}
\mathcal{B}(\chi_{c2}\to p \bar p K^{0}_{S} K^{-} \pi^{+} + c.c.)=(5.63\pm0.28\pm0.46)\times10^{-5}.
\nonumber
\end{equation}

We have also found evidence for $\chi_{c0,1,2} \to \bar{p} \Lambda(1520) K^0_S \pi^{+} + c.c.$ with statistical significances of 3.3$\sigma$, 5.7$\sigma$ and 7.0$\sigma$, respectively.
The branching fractions of these decays are determined to be
\begin{equation}
	\mathcal{B}(\chi_{c0}\to \bar p \Lambda(1520) K^{0}_{S} \pi^{+}+c.c)=(1.61^{+0.68}_{-0.64}\pm0.23)\times10^{-5},
	\nonumber
\end{equation}
\begin{equation}
	\mathcal{B}(\chi_{c1}\to \bar p \Lambda(1520) K^0_S \pi^{+}+c.c)=(4.06^{+0.80}_{-0.76}\pm0.52)\times10^{-5},
	\nonumber
\end{equation}
\begin{equation}
	\mathcal{B}(\chi_{c2}\to \bar p \Lambda(1520) K^0_S \pi^{+}+c.c)=(4.09^{+0.87}_{-0.84}\pm0.42)\times10^{-5}.
	\nonumber
\end{equation}
~These measurements of $\chi_{cJ}$ decays to a $p\bar{p}$ pair and three mesons  will be helpful for understanding $\chi_{cJ}$ decay mechanisms. In addition, we have examined the $p\bar p$ invariant mass spectra for $\chi_{cJ}\to p \bar p K^{0}_{S} K^{\mp} \pi^{\pm}$, and
no significant enhancement around the $ p \bar p$ mass threshold is found.

\section{Acknowledgement}

The BESIII Collaboration thanks the staff of BEPCII and the IHEP computing center for their strong support. This work is supported in part by National Key R\&D Program of China under Contracts Nos. 2020YFA0406300, 2020YFA0406400, 2023YFA1606000; National Natural Science Foundation of China (NSFC) under Contracts Nos. 12035009, 11875170, 11635010, 11735014, 11935015, 11935016, 11935018, 12025502, 12035013, 12061131003, 12192260, 12192261, 12192262, 12192263, 12192264, 12192265, 12221005, 12225509, 12235017, 12361141819; the Chinese Academy of Sciences (CAS) Large-Scale Scientific Facility Program; the CAS Center for Excellence in Particle Physics (CCEPP); Joint Large-Scale Scientific Facility Funds of the NSFC and CAS under Contract No. U1832207; 100 Talents Program of CAS; The Institute of Nuclear and Particle Physics (INPAC) and Shanghai Key Laboratory for Particle Physics and Cosmology; German Research Foundation DFG under Contracts Nos. FOR5327, GRK 2149; Istituto Nazionale di Fisica Nucleare, Italy; Knut and Alice Wallenberg Foundation under Contracts Nos. 2021.0174, 2021.0299; Ministry of Development of Turkey under Contract No. DPT2006K-120470; National Research Foundation of Korea under Contract No. NRF-2022R1A2C1092335; National Science and Technology fund of Mongolia; National Science Research and Innovation Fund (NSRF) via the Program Management Unit for Human Resources \& Institutional Development, Research and Innovation of Thailand under Contracts Nos. B16F640076, B50G670107; Polish National Science Centre under Contract No. 2019/35/O/ST2/02907; Swedish Research Council under Contract No. 2019.04595; The Swedish Foundation for International Cooperation in Research and Higher Education under Contract No. CH2018-7756; U. S. Department of Energy under Contract No. DE-FG02-05ER41374.

\end{document}